\documentclass[atmp]{ipart_v1} %%line 721 ,|line 989 \,|
\firstpage{1249}
\Year{2014}
\Vol{18}
\Issue{6}

\newcommand{\beq}{\begin{equation}}
\newcommand{\eeq}{\end{equation}}
\newcommand{\beqa}{\begin{eqnarray}}
\newcommand{\eeqa}{\end{eqnarray}}
\newcommand{\nn}{\nonumber}

\newcommand{\half}{\frac{1}{2}}

\newcommand{\we}{\wedge}

\newcommand{\der}{\partial}

\newcommand{\ka}{\varkappa}

\newcommand{\Psib}{\overline{\Psi}}
\newcommand{\Phib}{\overline{\Phi}}

\newcommand{\what}[1]{\widehat{#1}}

\newcommand{\bx}{{\bf x}}

\newcommand{\BPsi}{{\bf \Psi}} 

\newcommand{\BXi}{{\bf \Xi}}

\newcommand{\BN}{{\bf N}} 
 
\numberwithin{equation}{section}

\begin{document}

\author[Igor V. Kanatchikov]{Igor V. Kanatchikov\vskip1em  {(\it In loving memory of Pavel Efimov)}}

\title[Precanonical quantization and $\cdots$]{Precanonical quantization and the Schr\"odinger wave functional revisited}

\begin{abstract}
We address the  issue 
of the relation between the canonical functional Schr\"odinger  representation 
in quantum field theory and the approach of 
precanonical field quantization proposed by the author, 
which requires neither a distinguished time variable 
nor infinite-dimensional spaces of field configurations. 
We argue that the standard functional derivative Schr\"odinger equation 
can be derived  from the precanonical Dirac-like covariant generalization 
of the Schr\"odinger equation under the  formal limiting transition  $\gamma^0\varkappa\rightarrow\delta(\mbox{\bf 0})$, 
where the constant $\varkappa$ naturally 
appears within the precanonical quantization 
 %on dimensional grounds 
as the inverse of a small ``elementary volume" of space.   
We obtain a formal explicit expression of the Schr\"odinger 
wave functional as a continuous product 
of the Dirac algebra valued precanonical wave functions, 
which are defined 
on the finite-dimensional covariant configuration 
space of the field variables and space-time variables. 
\end{abstract} 

\maketitle

\section{Introduction }
                    
Precanonical quantization of field theory was proposed in 
\cite{qs96,ikanat2,ikanat3} as 
an analogue of canonical quantization which does not distinguish between 
the %?? 
space and time variables, and hence is more compliant with the 
%relativity theory. 
concept of relativistic space-time. 
It is inspired by the De Donder-Weyl (DW) analogue  
of the Hamiltonian formulation known in the calculus of variations 
\cite{dw,w35,rund,ka}.  %%?
%which 
This formulation 
can be viewed as a generalization of the Hamiltonian description 
to field theory, %%? 
 such that all space-time variables are treated on 
 an equal footing 
as analogues of the  time parameter in mechanics. 
 %and hence it allows us to view 
 %thus allowing us to view 
%This 
Such an approach allows us to view 
fields as finite-dimensional systems {\em changing } 
in space-time rather than infinite-dimensional mechanical 
systems {\em evolving } in time. 

Specifically, given a first order Lagrangian function  
$L(y^a, y^a_\mu, x^\nu)$ on the space of  field variables $y^a$, 
their first space-time derivatives $y^a_\mu$,  and  the space-time 
variables $x^\mu$,  one can introduce 
the %? 
Ham\-il\-tonian-like variables:   
 $p_a^\mu\linebreak :=\der L / \der y^a_\mu$ ({\em polymomenta})  
 \mbox{\rm and} 
  $H=H(y^a,p^\mu_a,x^\nu):=y^a_\mu p^\mu_a -L$    
 ({\em DW Hamiltonian function}),  
and write the Euler-Lagrange  field equations in the\linebreak 
 De Donder-Weyl Hamiltonian-like form  \cite{dw,w35,rund,ka}: 
\beq
\der_\mu y^a (x) = \frac{\der H}{\der p^\mu_a}~, 
\quad 
\der_\mu  p^\mu_a (x) =- \frac{\der H}{\der y^a }~, 
\eeq 
 %if 
provided 
$\det \left ( \frac{\der^2 L}{\der y^a_\mu \der y^b_\nu} \right ) 
\neq 0$. 

In this formulation the analogue of the extended configuration space is 
the finite-dimensional space of field variables $y^a$ and space-time 
variables $x^\mu$, and the analogue of the extended phase space is the 
polymomentum phase space of variables $(y^a, p_a^\mu, x^\mu)$. Note that 
when the number of space-time dimensions 
$n=1$, %comma?  
this formulation reproduces the standard Hamiltonian formulation of mechanics.  
At the same time,  it provides an alternative to the standard extension of 
the Hamiltonian formulation to field theory, the canonical formalism,   
 which is 
 %???is it clear that WHICH refers to "the standard extension"
based on the space-time 
decomposition and infinite-dimensional spaces of field configurations  
and, therefore, is 
 %limited to the globally hyperbolic space-times only.  
 applicable only on globally hyperbolic space-times. 

%The term ``precanonical'' stems from 
We will refer to the DW Hamiltonian formalism and the related quantization 
as ``precanonical'' based on 
the observation that 
the symplectic structure and other structures of the 
canonical Hamiltonian formalism in field theory 
can be related to, or derived from those of 
the De Donder-Weyl (or polysymplectic, or multisymplectic) 
formalism by restricting the latter to the surface of initial field 
configurations at a fixed moment of time 
(which we call the Cauchy surface) 
and integrating over it 
\cite{ikanat,kijtul,sniat,gotay2,gimmsy,ikan-pla,helein2}.

Though the idea of an approach to field quantization based on 
the DW Hamiltonian formulation  was discussed 
 already by Hermann Weyl himself in 1934 \cite{w34}, it was 
 not further developed, %comma? 
because an analogue of the Poisson bracket in 
the DW Hamiltonian formalism was not known until it was constructed 
using the polysymplectic structure on the polymomentum phase space 
in \cite{ikanat,bial96,go96} 
(see also 
 \cite{paufler,forger,helein,mydirac,roy,baez,vanker} 
for later discussions and generalizations). 
 
Since the Poisson bracket in precanonical DW formulation 
\cite{ikanat,bial96,go96} 
is defined on 
differential forms representing the dynamical variables of field theory,   
and it leads to the Gerstenhaber algebra structure 
%(%the?
%i.e. graded  Lie bracket + Grassmann product with the respective grades 
%differing by $1$) 
generalizing the 
standard Poisson algebra structure, its quantization has been 
 shown to lead 
to a quantum formalism with  the space-time Clifford algebra valued operators 
and wave functions,    
 %comma here?  
and the following precanonical 
 generalization of the Schr\"odinger equation \cite{qs96,ikanat2,ikanat3}: 
\beq \label{nse}
i\hbar \ka \gamma^\mu\der_\mu \Psi = \what{H}\Psi, 
\eeq 
where $\Psi=\Psi(y^a, x^\mu)$ is a Clifford-valued wave function on 
the covariant configuration space, 
 $\what{H}$ is the 
 %(partial differential) 
operator of the DW Hamiltonian function, 
which contains the partial derivatives with respect to the field variables: $\der_{y^a}$,  
and 
$\varkappa$ is a (large) constant of the dimension 
 {\tt length}$^{-{(n-1)}}$ in $n$ space-time dimensions, %%? 
which appears on dimensional grounds when, e.g.,  the 
differential 
 %>>exterior 
form corresponding to the infinitesimal volume element of space: 
$d\bx :=dx^1\we\cdots\we dx^{n-1}$,  is mapped to the  
 %proper 
corresponding element of the 
space-time Clifford (or Dirac) algebra %, viz.,  
 using the ``quantization map'' $q$ known in the theory of Clifford algebras 
 (see e.g. \cite{mein-cl}):   
 %represented as 
\beq \label{whatdx} 
d\bx\xrightarrow{q}%\what{d\bx}=
\frac{1}{\ka}\,\gamma_0 . 
\eeq 
%using the ``quantization map'' $q$ known in the theory of Clifford algebras 
%(see e.g. \cite{mein-cl}). 

 In a sense, the procedure  of precanonical quantization 
 effectively introduces in the theory an 
 ``elementary volume'' scale $1/\ka$ 
without any a priori assumptions regarding 
the microscopic structure of space-time. 
% (similar to $\hbar$ introducing the scale of 
%the minimal action in classical phase space).
The limit of the vanishing ``elementary volume'' scale 
corresponds to $1/\ka\rightarrow 0$ or, more precisely, 
$$
\frac{1}{\ka}\,\gamma_0 \xrightarrow{q^{-1}}  d\bx  
$$
or, equivalently,  
\beq \label{whatdx2} 
\gamma^0\ka\xrightarrow{q^{-1}}\delta^{n-1}(\mbox{\bf 0}) ,  
\eeq 
where  $q^{-1}$ is the inverse of the Clifford algebraic ``quantization map'' 
from the Grassmann algebra  of forms to the Clifford algebra of Dirac matrices. 

The precanonical  wave function  $\Psi(y, x)$ can be 
interpreted as the probability amplitude of finding the value of the field 
in the interval $[y,y+dy]$ when observed in the vicinity of the space-time 
point 
$[x,x+dx]$. This interpretation is supported by the conservation law 
$$
\der_\mu \int \! dy \, {\sf Tr}(\Psib \gamma^\mu \Psi) = 0 ,  
 $$ 
which follows from (\ref{nse}) 
and the self-adjointness of $\what{H}$ 
with respect to the indefinite scalar product  $\int\! dy {\sf Tr}(\Phib  \Psi)$, 
$\Psib := \gamma^0 \Psi^\dagger \gamma^0$, 
and the positive definiteness of  the Frobenius norm 
 % \Psib \gamma^0 \Psi. 
 ${\sf Tr}(\Psi^\dagger \Psi)$.  
%former text: 
%if $\what{H}$ is self-adjoint 
%with respect to the scalar product  $\int\! dy {\sf Tr}(\Phib  \Psi)$. 

Note that at $n=1$ the 
formulation of precanonical quantization 
essentially reduces to the conventional quantum mechanics formulated in terms 
of complex wave functions. 

The description of quantum fields based on precanonical quantization 
 %is, however, 
appears to be 
fundamentally different from the familiar formulations of quantum field theory. 
By abandoning 
  %avoiding 
the usual explicit treatment of fields as infinite-dimensional Hamiltonian systems 
  %{\em evolving} in time 
and replacing the starting point of quantization with 
the %covariant 
De Donder-Weyl Hamiltonian formalism 
  %which treats fields rather as systems {\em changing} in space-time, 
%%we have lost 
 we lose 
 an obvious connection 
 %is lost %correct grammar? good style? order of words??? 
with the concepts of the standard formulations of QFT, 
such as free particles, which straightforwardly follow from the conventional 
treatment and are crucial for the comparison of the results of quantum 
field theory with the experiments, at least in the perturbative regime. 
On the other hand, the construction of precanonical quantization is 
non-perturbative, explicitly compliant with the relativistic 
nature of space-time, 
 %principles,  
and it seems to be potentially better or 
 easier 
defined mathematically than the infinite-dimensional constructions of QFT.  

In spite of the fact that there already have been 
attempts to apply precanonical quantization  
in quantum Yang-Mills theory 
\cite{ikan-ym}, quantum gravity \cite{ik-grav,ik3,ik4,ik5,rov,oriti}
and string theory \cite{castro-string},   
the lack of good understanding of the  connections of 
the description suggested by 
precanonical quantization with the concepts of standard QFT 
has been hindering  
 %its 
 %%it is clear that ITS means precanonical quantization?? 
applications  
 of precanonical quantization 
and its further development 
 so far.  
 
For the sake of completeness let us quote  several other attempts  
of  field quantization inspired by 
the %%? 
covariant Hamiltonian-like  formulations in the calculus of variations:  
 %can be found in 
 %have appeared in the recent literature, 
see e.g. 
\cite{hamburg,sardan-q,stoyan,rov-q,nik-q,harrivel}. 

In our previous papers \cite{ikan-pla,ikan-ym} we tried to understand 
the connection of precanonical quantization 
 and its description of 
quantum fields in terms of Clifford-valued wave 
 {\em functions } $\Psi(y,x)$ 
 %%on the finite dimensional analogue of extended configuration space, 
 %%viz. the space of 
 %%field variables $y$ and space-time variables $x^\mu= (\bx,t)$, 
with the standard canonical quantization in the functional 
Schr\"odinger  representation \cite{hatfield,schweber,symanzik,jackiw,corichi},  
 %comma?  
where the states of quantum fields are described by the wave 
{\em functionals } $\BPsi([y(\bx)],t)$
on the infinite-dimensional configuration space of field configurations 
$y(\bx)$ at the instant of time $t$. 
  However, the discussion 
in \cite{qs96,ikan-pla,ikan-ym} 
has not established a convincing connection because of 
the following shortcomings of the arguments: 

(i) the simplifying ultra-locality assumption that the 
Schr\"odinger functional 
 $\BPsi([y(\bx)],t)$ 
 can be represented  in the form of the 
 %continual >> previously used term!!! 
 continuous 
product of precanonical wave functions $\Psi(y,\bx,t)$ over all points of 
 %space $\bx$ 
 the surface $y=y(\bx)$ (cf. Eq. (\ref{ul}) below) 
neglects the  correlations between the field values at %in 
the space-like separated points, thus  contradicting 
both the known explicit form of the exact solutions of the 
functional Schr\"odinger equation for free field theories 
\cite{hatfield,schweber,jackiw}
and the known behaviour of the Wightman functions of free fields;  

(ii) an attempt to take  those correlations into account by means of 
a functional ``unitary transformation'',   which would transform the 
equation in functional derivatives satisfied by the ultralocal 
 %continual 
continuous product Ansatz 
 %$\BPsi \sim \prod_\bx \Psi ... $ 
of~\cite{ikan-pla} 
into the equation for the wave functional in the Schr\"odinger picture,  
leads to a non-integrable equation in functional derivatives 
for the functional operator $e^{i\BN}$ determining such a transformation:  
$\delta \BN/\delta y(\bx) = \gamma^i\der_i y(\bx); $

(iii) the representation of the continuous product 
$\prod_\bx \Psi(\bx)$ 
 %as 
 %$$\exp{\left (\varkappa\int d\bx \ln(\Psi(\bx))\right )}$$  
in the form  $e^{{\varkappa\int d\bx \ln(\Psi(\bx))}}$ 
used in \cite{ikan-pla} 
is questionable for Clifford-valued $\Psi(\bx)$, which may not 
commute at different points $\bx$; besides,   
its functional differentiation implies 
the existence of the inverse 
$\Psi^{-1}(\bx)$ for all $\bx$, which is too restrictive and 
even impossible to define %if $\Psi$ is spinor-valued. 
for arbitrary Clifford algebra elements $\Psi$. 
 %----$e^W$---- 

In this paper, we revisit our  earlier 
treatment \cite{ikan-pla}  of the relation between 
the precanonical wave function and the Schr\"odinger wave functional 
using the example of scalar field theory. 
 We follow the conventions and notations of that paper, 
%as well as we 
and we also 
refer to it both for a brief outline of the elements of precanonical 
and canonical quantization of scalar field theory and 
an explanation of some constructions to be used here. It will be shown below  
that a minimal 
%modification 
generalization of the ultra-local Ansatz 
used in \cite{qs96,ikan-pla,ikan-ym} allows us to take 
into account  the space-like correlations, %or non-ultralocality, 
which were neglected in the earlier treatment  of \cite{ikan-pla}, 
 and to derive a formula expressing the Schr\"odinger 
wave functional in terms of 
 %the?? 
precanonical wave functions. 

\section{Precanonical wave functions and 
the Schr\"odinger\\ wave functional} 

We restrict ourselves to the example of the scalar field 
theory given by 
\beq L=\half \der_\mu y \der^\mu y - V(y), \nn \eeq
where the potential term $V(y)$ also includes the mass term $\half m^2y^2$ 
and $\hbar=1$ henceforth. 

The Schr\"odinger wave functional of the quantum scalar field 
 %is known to obey 
 obeys the Schr\"odin\-ger equation in functional derivatives 
\cite{hatfield,schweber,jackiw,symanzik,corichi}: 
\beq \label{schr-func}
i\der_t \BPsi = \int d\bx \left \{ -\frac{1}{2}\frac{\delta^2}{\delta y(\bx)^2} 
+ \half (\nabla y(\bx))^2 + V(y(\bx))
\right \} \BPsi .
\eeq

Our task is to clarify how the description of quantum fields 
 in terms of the wave functional 
$\BPsi([y(\bx)],t)$ is related to the description in terms of the precanonical 
wave function $\Psi(y,x)$, and how the Schr\"odinger equation 
of the canonical quantization approach, Eq. (\ref{schr-func}), is related to 
the Dirac-like partial differential equation, Eq. (\ref{nse}), playing the role of the Schr\"odinger 
equation 
 %for the  precanonical wave function.
in the precanonical quantization approach.

Note that if the precanonical wave function $\Psi(y,x)$ 
 %of the precanonical quantization approach 
has the  probabilistic interpretation 
as the probability amplitude of finding the value $y$ of the field 
 at the 
space-time point $x$, then the Schr\"odinger wave functional 
$\BPsi([y(\bx)],t)$, which is the probability amplitude of observing 
the field configuration $y(\bx)$ on the space-like hypersurface of constant 
time $t$, should be given by a certain composition of precanonical amplitudes 
$\Psi(y,\bx,t)$ taken along the Cauchy surface $\Sigma: (t=$const, $y=y(\bx)) $ 
in the covariant configuration space 
of variables $(y,x)$. 
  %given by: $t=$const, $y=y(\bx)$. 

If we assume that the probability amplitudes of observing 
the %? 
field values 
$y$ are independent in space-like separated points, 
then $\BPsi([y(\bx)],t)$ 
 %could be 
 is given by the product over all points of $\Sigma$ of the wave function 
$\Psi(y,x)$ restricted to $\Sigma$: 
$\Psi(y,x)|_\Sigma=\Psi_\Sigma(y=y(\bx),\bx,t)$,
 i.e.  
\beq \label{ul}
 \BPsi \sim \prod_{\bx\in\Sigma}\Psi_\Sigma(y=y(\bx),\bx,t) .
\eeq
This ultra-locality  
assumption is, however, unphysical and 
 an improved 
representation of the Schr\"odinger wave functional in terms of
$\Psi(y,x)|_\Sigma$ 
 has to be found, which would take into 
account  the correlations of  
 the amplitudes $\Psi(y,x)$ 
 %in? 
 at space-like separated points. 

The task is similar to the probability theory,  where the joint probability of 
two events $A$ and $B$ is given in general by $P(A,B) = P(A|B)P(B)$, 
where  $P(A|B)$ is the conditional probability of $A$ given $B$, 
 %comma? 
which reduces to  $P(A)$ only if the events $A$ and $B$ are independent.
 In our case we have a 
continuum of events of obtaining the values $y_\bx$ of the field 
in %at?? 
the corresponding points $\bx$ of the hypersurface of constant time $t$, %%comma?
and their respective probability amplitudes $\Psi(y_\bx,\bx,t)$ given by 
the precanonical wave function taken along the surface $\Sigma$: 
$\Psi(y=y(\bx),\bx,t)$. 

As a minimal deviation from the simplest 
ultralocal product formula 
 %$\BPsi \sim \prod_{\bx\in\Sigma}\Psi(y,x)|_\Sigma$  
 %above 
 in (\ref{ul}) 
let us assume that the correlations between space-like separated points can 
be taken into account by 
 %transformation 
a multiplication of $\Psi(y,x)|_\Sigma$ by some function of the field 
configuration denoted $U(y(\bx))$, so that the Schr\"odinger  functional 
can be given by a modified product formula 
$\BPsi \sim \prod_{\bx\in\Sigma}U(y(\bx)) \Psi(y,x)|_\Sigma$. 
 %In what follows we will 
 Below we 
  %investigate if 
  show that this minimal 
   %modification 
   generalization of the ultra-local 
product formula 
of \cite{ikan-pla} is general enough to formulate a connection between 
the precanonical description of quantum fields and the canonical description 
in the functional Schr\"odinger representation. 

 Thus, let us 
assume that the Schr\"odinger wave functional has the form 
\beq \label{assume}
\BPsi =  {\sf Tr}\left \{ 
\prod_\bx U(y(\bx)) 
\Psi_\Sigma (y(\bx), \bx, t)
\right \} ,
\eeq
where $U(y(\bx))$ is a matrix transformation which, 
  in order to go beyond the ultra-locality assumption in (\ref{ul}), 
%depends on 
is supposed to depend on 
 %the configuration 
the value of the field $y(\bx)$  and  
 %perhaps 
its derivatives at the point $\bx$.  
    %(in order to go beyond the ultra-locality assumption in (\ref{ul})). 
 %%is "in order" superficial here??? 
  %REMARK: first derivative <> bilocality??? 
  
The latter expression means that 
 for any  $\bx$  the functional 
$\BPsi$  can be written as 
 %follows: 
 \beq \label{psiphi}
\BPsi =  {\sf Tr}\left \{ 
\BXi(\breve{\bx},t ) 
U(y(\bx))
\Psi_\Sigma (y(\bx), \bx, t)
\right \},
\eeq 
%%for any  $\bx$, 
where 
\beq
\BXi(\breve{\bx},t)  : =  
\prod_{\bx' \neq \bx} U(y(\bx')) 
\Psi_\Sigma (y(\bx'), \bx', t) 
\eeq 
and the 
continuous product here implies 
 %a symmetrization 
 the cyclic permutations 
over all points $\bx'$. 
 
This observation facilitates the calculation of functional derivatives 
of $\BPsi$, viz.,  
\beq
\frac{\delta \BPsi }{\delta y(\bx)} = 
{\sf Tr}\left \{
\BXi (\breve{\bx}) \frac{\delta U (\bx) }{\delta y(\bx)} \Psi_\Sigma(\bx) 
+ \BXi (\breve{\bx}) U (\bx) \delta(\mbox{\bf 0}) \der_y \Psi_\Sigma(\bx) 
\right \}  
\eeq 
and 
\begin{align}
    \frac{\delta^2 \BPsi }{\delta y(\bx)^2} &= 
{\sf Tr}\left \{
 \BXi (\breve{\bx}) \frac{\delta^2 U (\bx) }{\delta y(\bx)^2} \Psi_\Sigma(\bx) 
+ \BXi (\breve{\bx}) U (\bx) \delta(\mbox{\bf 0})^2 \der_{yy} \Psi_\Sigma(\bx)
\right .\label{delta2}\\
&\quad \left. 
 +2\, \BXi (\breve{\bx}) \frac{\delta U (\bx) }{\delta y(\bx)} \delta(\mbox{\bf 0}) \der_y \Psi_\Sigma(\bx) 
\right \} ,\notag 
\end{align} 
where the shorthand notations  
$\BXi(\breve{\bx})$ for  $\BXi(\breve{\bx},t)$, $U(\bx)$ for  $U(y(\bx))$,  
and $ \Psi_\Sigma(\bx)$ for $\Psi_\Sigma (y(\bx), \bx, t)$
are introduced. Note that 
the %??
 $(n-1)$-dimensional $\delta(\mbox{\bf 0})$ appears here as a 
result of 
%the ?? 
functional  differentiation of a function 
 %in some point 
with respect to itself 
 %in 
   at the same spatial point.

The time derivative of $\BPsi$ in (\ref{assume}) 
 %with respect to the time variable 
 is given by the chain rule:  
\beq \label{dete}
\der_t \BPsi = {\sf Tr}\left \{ 
\int %_\Sigma d\bsigma(\bx) 
d\bx~ %\sqrt{g_\Sigma(\bx)}\ 
 \frac{\delta \BPsi }{\delta \Psi^T_\Sigma(\bx) } 
% \beta ??
\der_t \Psi_\Sigma(\bx) 
\right \} ,
\eeq 
 where $ \Psi^T$ denotes the transpose of  $\Psi$. 
Using (\ref{psiphi}) we obtain  
\beq \label{detepsi}
i\der_t \BPsi = {\sf Tr}\left \{ 
\int d\bx~ %\sqrt{g_\Sigma(\bx)}\ 
 \BXi(\breve{\bx}) U (\bx) 
\delta(\mbox{\bf 0})
  i\der_t \Psi_\Sigma(\bx) 
\right \} .
\eeq
Hence,  the time evolution of the wave functional $ \BPsi$ 
is totally dictated by the time 
evolution of the precanonical wave function  restricted to the Cauchy surface, 
$\Psi_\Sigma(\bx)$. 

The time evolution of the restricted wave function $\Psi_\Sigma(\bx)$ 
 is given by our Dirac-like precanonical Schr\"odinger equation 
on $\Psi (y,x)$,  Eq. (\ref{nse}),  
 restricted to the Cauchy surface $\Sigma$  
  (cf. \cite{ikan-pla}): 
\beq \label{nsesig}
i \der_t \Psi_\Sigma(\bx) = 
-i \alpha^i\frac{d}{dx^i} \Psi_\Sigma(\bx)  
+i \alpha^i\der_i y(\bx) \der_{y} \Psi_\Sigma(\bx) + 
\frac{1}{\ka}\beta(\what{H} \Psi)_\Sigma(\bx) .
\eeq 
Here  $\beta:=\gamma^0$, $\alpha^i:=\beta\gamma^i$,  
$\frac{d}{dx^i}$ denotes the total derivative along $\Sigma$ in jet space 
$(y, y_i, y_{ij},...)$, such that $y_\Sigma = y(\bx)$, $y_i{}_\Sigma = \der_i y(\bx)$, 
$y_{ij}{}_\Sigma = \der_{ij} y(\bx)$, etc.: 
\beq \label{total}
\frac{d}{dx^i}:=\der_i+\der_iy(\bx)\der_y+\der_{ij}y(\bx)\der_{y_j} +\cdots,
\eeq
and,  in the specific case of 
the %%?? 
scalar field $y$ (cf. \cite{qs96,ikanat2,ikanat3,ikan-pla}), 
\beq \label{hscalar}
(\what{H} \Psi)_\Sigma(\bx) 
= -\half\ka^2\der_{yy}\Psi_\Sigma(\bx)  +V(y(\bx))\Psi_\Sigma(\bx)~. 
\eeq 
 %(cf. \cite{qs96,ikanat2,ikanat3,ikan-pla}).  

It is easy to see that if 
 equation (\ref{nsesig}) is  substituted in (\ref{detepsi}),  
the potential term in (\ref{hscalar}) yields 
$$ 
{\sf Tr}\left \{ 
\int\! d\bx~ 
\BXi(\breve{\bx}) U (\bx) 
\delta(\mbox{\bf 0})
  \frac{1}{\ka}\beta V(y(\bx))\Psi_\Sigma(\bx) 
\right \} .
$$
 It will reduce to the potential term in the 
functional Schr\"odinger equation:  
$$
\int\! d\bx\, V(y(\bx)) \BPsi, 
$$
with $\BPsi$ given by (\ref{psiphi}), %%%comma 
if, in some mathematical sense,  $\ka\beta$ is replaced by, or 
  goes over into  $\delta(\mbox{\bf 0})$: 
\beq \label{kadelta}
\ka\beta\rightarrow \delta(\mbox{\bf 0}) . 
\eeq 

We also notice that, quite remarkably,  under the same condition  (\ref{kadelta}) 
the term  $\ka^2 \der_{yy}\Psi_\Sigma$ in (\ref{hscalar}) 
 %goes over into 
 reproduces  
the second term in the second functional derivative 
of $\BPsi$ in Eq. (\ref{delta2}). 

 Thus,  the %relation 
 condition (\ref{kadelta}) establishes a 
 %condition 
 formal limiting map under which 
the transition from precanonical to the functional Schr\"odinger description 
is possible. As it was explained in Eq. (\ref{whatdx2}), 
%it 
this map coincides with the inverse quantization map $q$ 
in the limit of the vanishing elementary volume  $1/\varkappa \rightarrow 0$. 
%We note that it coincides with the inverse quantization map 
%and the limit of the vanishing elementary volume  in Eq. (\ref{whatdx2}).  

Next, we note that in order to obtain a description in terms of the 
wave functional $\BPsi$ alone, without any reference to precanonical wave functions,  
 the remaining terms in front 
of $\der_y \Psi_\Sigma$ in (\ref{nsesig})  and (\ref{delta2}) 
should cancel each other, at least in the limiting case (\ref{kadelta}). 
  This requirement  leads to the equation on $U(\bx)$:
 %condition 
\beq \label{u-eqn1}
%\sqrt{g_\Sigma(\bx)}\ 
U(\bx) i \beta \gamma^i \der_i y(\bx)  
+ \frac{\delta U(\bx)}{\delta y(\bx)} =0  , 
\eeq 
 %which determines $U(\bx)$.  This equation 
 which is not integrable. 
 %(cf. e.g. \cite{zador}). 
 However, 
by taking into account the condition (\ref{kadelta}), 
in the corresponding limit we can re-write (\ref{u-eqn1}) as  
\beq \label{u-eqn}
U(\bx) i \gamma^i \der_i y(\bx)  \delta(\mbox{\bf 0})
+ \ka \frac{\delta U(\bx)}{\delta y(\bx)} =0 , 
\eeq
 %The solution of this equation 
 whose solution can be written in the form 
\beq \label{u}
U(\bx)= e^{-iy(\bx)\gamma^i\der_iy(\bx)/\ka} 
 \eeq 
up to 
 %an integrating 
 a factor, which can be implicitly taken into account 
in a redefinition of  $\Xi(\breve{\bx})$.   
Besides, from (\ref{u-eqn}), 
 %in the limit  
 under the limiting map (\ref{kadelta}), 
 %commas???  
it also follows that 
\beq 
 \frac{\delta^2 U (\bx) }{\delta y(\bx)^2}
= (\nabla y(\bx))^2 U (\bx) . 
\eeq 
Hence, in the limiting case (\ref{kadelta}) 
the first term in (\ref{delta2}) 
correctly reproduces the  term $\int\! d\bx (\nabla y(\bx))^2\BPsi$ 
in the functional derivative Schr\"odinger equation, Eq. (\ref{schr-func}). 

The last term to be considered is the total derivative term in (\ref{nsesig}).  
When inserted in (\ref{dete}) with the condition (\ref{kadelta}) taken into 
account,  %%commas? 
and then  integrated by parts,   it yields 
   a term proportional to 
 $${\sf Tr}\left \{ 
\int\! d\bx~ 
 \BXi(\breve{\bx}) 
\frac{d}{dx^i} U (\bx)\gamma^i
 \Psi_\Sigma(\bx)  
\right \} . 
 $$ 
Using the explicit form of $U (\bx)$ in (\ref{u}) 
to evaluate $\frac{d}{dx^i} U (\bx)$  %comma? 
and recalling 
the representation of $\BPsi$ in (\ref{psiphi}), this term 
is transformed to the form 
$$ %\sim 
\int\! d\bx \left ( y(\bx)\gamma^i\gamma^j \der_{ij}y(\bx)
+ \gamma^i\der_{i}y(\bx)\gamma^j\der_{j}y(\bx)\right ) \BPsi 
 , %comma? 
$$
which obviously vanishes upon integration by parts. 
Consequently, the total derivative term in 
(\ref{nsesig}) does not contribute to the functional 
derivative Schr\"odinger equation on $\BPsi$. 

Thus, we have demonstrated that the substitution of the precanonical 
Schr\"o\-din\-ger equation restricted to the Cauchy surface $\Sigma$, 
Eqs. (\ref{nsesig}--\ref{hscalar}), into the 
expression 
(\ref{dete}) for the time derivative of the functional (\ref{assume}) 
constructed from precanonical wave functions 
 allows us to  
  %obtain 
  reproduce, in the limiting case  (\ref{kadelta}), 
  all the 
  terms which are present in the 
canonical functional derivative Schr\"odinger equation 
(\ref{schr-func})  and to cancel those 
 which are 
 %missing 
 absent 
 there. 
The procedure fixes both the condition (\ref{kadelta}), %%comma? 
under which the transition from 
 the precanonical description 
   %an implicit word "description is omited - correctly? 
 to the canonical   description in terms of a Schr\"odinger wave functional 
  is possible,  %%comma?
and  the transformation $U$ in  (\ref{assume}),   
 %which then correctly reproduces 
 which allows us to reproduce the term $\half (\nabla y(\bx))^2$ %term 
in the canonical Hamiltonian in (\ref{schr-func}). 
This is the latter  term which is responsible for the non-ultralocal 
behaviour of quantum fields (see e.g. \cite{klauder}), 
whose natural derivation 
was problematic  in our previous paper \cite{ikan-pla}. 

 Moreover, the above consideration %procedure  
 also yields an explicit 
 limiting expression of the 
Schr\"odinger wave functional as the  continuous product 
over all spatial points 
%of the space  
of $U$-transformed precanonical wave functions restricted 
to the Cauchy surface $\Sigma$.  %, viz.,   
Namely, using (\ref{u}) in (\ref{assume}),  we obtain: 
\beq \label{theformula}
 \BPsi([y(\bx)],t) = {\sf Tr}\left \{ 
\prod_\bx  e^{-iy(\bx)\gamma^i\der_iy(\bx)/\ka}
\Psi_\Sigma (y(\bx), \bx, t)
\right \}
%_{\mbox{\large $\rvert$} \scriptscriptstyle    %\tiny 
% \frac{1}{\ka} \beta \mapsto d\bx } 
%%|_{\beta\ka\rightarrow \delta(\mbox{\bf 0})} ~ 
.
\eeq 
%which 
This formula is valid %under the limiting condition  
  %in combination with the inverse Clifford algebraic quantization map 
in the limit of an infinitesimal ``elementary volume'' $1/\ka$ 
 %and when 
%%combined with  
 and under the formal limiting map $\ka\beta\xrightarrow{q^{-1}}\delta(\mbox{\bf 0})$ 
 in the expression under  the continuous product sign. 
 In the subsequent paper \cite{inprep} we have shown that in this limit the formula in (\ref{theformula}) can be transformed to the 
 multidimensional product integral invented by Volterra \cite{volterra} and used to construct the well-known  vacuum state wave functional of free scalar field theory from the 
 ground state solutions of the precanonical Schr\"odinger equation.    
%%ADD about the product integral??? 

\section{Conclusions}

 We have shown how  the canonical functional derivative 
Schr\"odinger equation (\ref{schr-func}) 
can be derived from the 
 partial differential covariant precanonical 
Schr\"odinger equation  (\ref{nse}) 
restricted to the Cauchy surface in the 
covariant configuration space of field theory.  
 We also obtained an explicit limiting 
expression of the Schr\"odinger wave functional 
 %of the scalar field theory 
 %of 
 from 
 the canonical quantization approach in terms of the 
precanonical wave functions defined on the finite dimensional space of 
field and space-time variables.  

Our result suggests that 
 %the?? 
quantum field theory 
 %resulting 
originating 
from the canonical quantization 
 %emerges as 
 can be viewed as 
 %the %a? 
 %``singular limit'' $\ka\rightarrow \delta(\mbox{\bf 0})$ 
 %limit 
 a limiting case $1/\ka \rightarrow 0$ 
of quantum field theory resulting from the precanonical 
quantization approach.  

Note that the result of our earlier discussion in \cite{ikan-pla} 
comes amazingly close to 
 %is amazingly similar to 
the formula (\ref{theformula}). 
However, there are important differences:

First, we found  here that the transformation 
which allows  us to fully take into account the deviations from 
ultra-locality  acts directly on precanonical wave functions rather 
than on the ultra-local functional constructed from them. This solves the 
problem with the non-integrability of the functional derivative 
equation for the corresponding transformation functional, which we 
tried to  circumvent in \cite{ikan-pla,ikan-ym}.
%%%acknowledge Zadorozhnii and his book here? 
 %%%
 
Second, we found that the Schr\"odinger functional description is 
possible only  under the limiting condition (\ref{kadelta}), %comma?  
 %$\beta\ka\rightarrow \delta(\mbox{\bf 0})}$  
which  essentially tells us that the description of quantum fields 
in terms of the Schr\"odinger wave functional corresponds to the limit 
of vanishing ``minimal volume'' $1/\ka$, or 
 $\ka \rightarrow \delta(\mbox{\bf 0})$, as it was already noticed 
in \cite{ikan-pla}. It should be noted  that the condition 
(\ref{kadelta}) complies both with the relativistic transformation laws 
of $\delta(\mbox{\bf 0})$ and the rules of precanonical quantization itself 
(cf. Eq.(\ref{whatdx})). Namely, (\ref{kadelta}) 
unifies two requirements to be 
fulfilled when a transition from 
 the precanonical description 
 to the functional Schr\"odinger 
description is being made: the absolute value of $\ka$ should tend to  
$\delta(\mbox{\bf 0})$ and the inverse of the ``quantization map''  
in (\ref{whatdx}) 
should be applied so that the Clifford-valued operator 
of the ``minimal volume'' $\beta/\ka$ 
 %goes over  into 
 is mapped (at  $\ka \rightarrow \delta(\mbox{\bf 0})$) 
 to the differential form $d\bx$ representing the 
classical infinitesimal volume element. 

Third, the identification of the condition when the transformation from 
precanonical to the functional Schr\"odinger representation is possible 
as the inverse quantization map in the limit $1/\ka\rightarrow 0$, Eq. 
(\ref{whatdx2}), has made superficial the use of  
the projector $\frac{1}{2}(1+\beta)$ introduced 
in the expression of the wave functional in  \cite{ikan-pla}.

%In mathematical terms, 
Mathematically speaking, by starting from  
 the assumption (\ref{assume}) 
and showing how it allows us to derive the canonical 
 functional derivative  Schr\"o\-dinger equation from the 
precanonical partial derivative Schr\"odinger equation,  
 %comma? 
and to fix the form of the transformation $U(y(\bx))$, 
we have proven that the Ansatz (\ref{assume}) is the sufficient condition %, 
which establishes a connection of  the Schr\"odinger wave functional 
with precanonical wave functions. 
In a forthcoming  paper \cite{inprep} we 
 have shown that this condition 
is also necessary. 
 
Note that,  although our result is obtained using the 
particular case of the scalar field theory, in the subsequent papers 
it will be demonstrated that it can be extended also to other fields, 
such as Yang-Mills and spinor fields. 

One should underline that the nature of the 
 ultra-violet parameter 
 %constant ? 
$\ka$ appearing in precanonical quantization is different 
from the ``minimal length'' scale 
introduced by hand in nonlocal field theories or 
field theories over  discrete/microstructured/non-commutative 
space-times. 
 %For example, 
 %it disappears 
\, This~\, parameter \,does not arbitrarily modify the relativistic space-time at small scales, 
and it was found to disappear from the final results for free field theories. 
We still have to investigate in detail what  the role of $\ka$ is  
in interacting field theories, renormalizable and non-renormalizable,  
and what is its possible role in the common renormalization techniques. 

In conclusion, let us recall that the main motivation of 
 %our developing of 
 the precanonical quantization approach has been its potential 
 %of being 
 to be a better synthesis of relativity and quantum theory 
 in the context of field theory, that could 
provide a better framework for quantization of gravity.
 % That is why 
We hope that the results of this paper can help to understand the 
physical content of precanonical quantization of gravity 
\cite{ik-grav,ik3,ik4,ik5,ik-tetrad} 
and, in particular, clarify its relation with the 
the canonical quantum gravity leading to the Wheeler-DeWitt equation, 
 %which is 
the ill defined quantum gravity analogue of the functional  Schr\"odinger equation, 
which could emerge as a singular limit of infinite $\varkappa$ 
from a better defined precanonical quantization of gravity.

%OLD:
%We hope to use the results of this paper 
 %may be used 
%for further development 
%of precanonical quantization of gravity \cite{ik-grav,ik3,ik4,ik5,ik-tetrad} 
%and, in particular, for better understanding of its relation 
 %to 
%with the canonical quantization 
%leading to the Wheeler-DeWitt equation, 
% the quantum gravity analogue of the functional  Schr\"odinger equation. 

\medskip 

{\bf Acknowledgements: }I would like to thank Aranya Bhattacherjee 
and Marek Czachor for their crucial support which has allowed me to 
complete this work.   % in spite of the circumstances. 
I also thank Alexander Stoyanovsky for sending to me a copy of 
his book \cite{stoyan}, Vladimir Zadorozhnii for his useful 
comments regarding the integrability of equations in variational derivatives,  
Andrew Forrester for his helpful remarks on the draft of the paper,  
and KCIK for its kind hospitality. 

%M. Zadorozhnii {\it Methods of variational analysis}, 
%R\&C Dynamics, Moscow-Izhevsk 2007 (in Russian). 
%ISBN 5-93972-489-2                            

\address{National Quantum Information Centre in Gda\'nsk (KCIK)\\
ul. W\l. Andersa 27, 81-824 Sopot, Poland\\
\email{kanattsi@gmail.com}}

\end{document}